# Realization of Flat Band with Possible Nontrivial Topology in Electronic Kagome Lattice

Zhi Li[1*], Jincheng Zhuang[1*], Lan Chen[2,3], Li Wang[1], Haifeng Feng[1,4], Xun Xu[1,4], Xiaolin Wang[1], Chao Zhang[1], Kehui Wu[2,3], Shi Xue Dou[1,4], Zhenpeng Hu[5,†], Yi Du[1,4†]

[1]*Institute for Superconducting and Electronic Materials (ISEM), Australian Institute for Innovative Materials (AIIM), UOW-Beihang Joint Research Centre, University of Wollongong, Wollongong, NSW 2525, Australia*
[2]*Institute of Physics, Chinese Academy of Sciences, Beijing 100190, P. R. China*
[3]*University of Chinese Academy of Sciences, Beijing 100049, P. R. China*
[4]*Beihang-UOW Joint Research Centre, Beihang University, Beijing 10191, P. R. China*
[5]*School of Physics, Nankai University, Tianjin 300071, P. R. China*
[†]To whom correspondence should be addressed: yi_du@uow.edu.au (Y. Du), zphu@nankai.edu.cn (Z. Hu)
*These authors contributed equally to this work.

## Abstract

The energy dispersion of fermions or bosons vanishes in momentum space if destructive quantum interference occurs in a frustrated Kagome lattice with only nearest-neighbour (NN) hopping. A discrete flat band (FB) without any dispersion is consequently formed, promising emergence of fractional quantum Hall states (FQHS) at high temperatures. Here, we report experimental realization of a FB with possible nontrivial topology in an electronic Kagome lattice on a twisted multilayer silicene. The electrons are localized in the Kagome lattice due to quantum destructive interference, and thus, their kinetic energy is quenched, which gives rise to a FB peak in density of states. A robust and pronounced one-dimensional edge state has been revealed at Kagome edge, which resides at higher energy than the FB. Our observations of the FB and the exotic edge state in electronic Kagome lattice open up the possibility towards the realization of fractional Chern insulators in two-dimensional materials.

**Main text**

In quantum mechanics, superposition of particle wave functions with unavoidable quantum interference gives birth to new quantum states. Quantum interference associated with frustrated geometry often gives rise to interesting strongly correlated phenomena and the emergence of fascinating nontrivial structures (*1-5*). Very recently, a number of works have proven theoretically that completely destructive quantum interference occurs if fermions or bosons are confined in a frustrated Kagome lattice with only nearest neighbour (NN) hopping, as illustrated in **Figure 1a** (*6-12*). Consequently, the dispersion relation of the trapped particles exhibits a completely flat characteristic, namely, a flat band (FB), that is, the energy is dispersionless in momentum space.

Inspired by these pioneering theoretical predictions, FBs of plasmons (*13*) and polaritons (*14*) have been consecutively realized in corresponding Kagome micro/nanostructures in recent years. In principle, an electronic FB can also be achieved in an electronic Kagome lattice. It is expected to be topologically nontrivial if the non-negligible spin-orbital coupling (SOC) effect is considered. Because the electronic FB mimics Landau levels but does not require a magnetic field, fractional quantum Hall states (FQHS) have been predicted to emerge when the FB is partially filled (*6-8*). Great efforts have been put in recent works in order to achieve this goal. Kagome lattices have been constructed artificially by assembling organic molecules (*15,16*), although the electronic FB was not observed due to negligible electron hopping between these organic molecules. Modulation of a two-dimensional (2D) electron gas by a local potential provides an alternative way to construct electronic lattices, such as the realization of a honeycomb electronic lattice in artificial molecular graphene (17). Indeed, the local potential of a nano-network of artificial quantum dots in a Kagome arrangement has successfully modulated a 2D electron gas and, consequently, created a new electronic band (*18*). Unfortunately, the local density of states (LDOS) of this band exhibited hexagonal periodicity rather than a Kagome arrangement, owing to quantum confinement. The band thus shows a dispersive nature and is not flat. So far, constructing electronic Kagome lattice and electronic FB remain experimental challenging, and have not been achieved as yet in any material systems.

Herein, we report the experimental realization of an electronic Kagome lattice with the emergence of an electronic FB on a twisted multilayered silicene (*19*, *20*) surface. The localized electronic states have been observed by scanning tunneling microscopy (STM). We used scanning tunneling spectroscopy (STS) to further reveal that this FB is induced by the electronic Kagome lattice and can induce exotic edge states even at 77 K. Our work provides an intriguing material candidate for the possible realization of FQHS.

The multilayer silicene film was prepared by layer-by-layer growth using molecular beam epitaxy, as described in our previous work (*21*). The sample surface exhibits a similar microscopic structure to that of the recently reported 2D multilayer silicene surface (*19*, *20*), which shows islands and domains several hundred nanometers in size. It was found that most of the surface area exhibits the $\sqrt{3} \times \sqrt{3}$ silicene honeycomb reconstruction with a lattice constant of 0.64 nm, as shown in **Figure 1b and 1d**. We also found that some islands (domains) display a larger periodicity, in which typical Kagome lattices with a lattice constant $a_{\text{Kagome}}$ of 1.7 nm were identified (**Figure 1c**). A comprehensive STM study (**see Supplementary Note 2**) on this Kagome lattice shows that it is induced by a composite moiré pattern due to an interlayer twisting of 21.8° between the silicene layers (*21*). At low sample biases ($|V_{\text{bias}}| < 0.2$ V), a $\sqrt{3} \times \sqrt{3}$ silicene structure with a faint height modulation can be simultaneously revealed in the Kagome lattice (**see Supplementary Note 3**). This suggests that the observed Kagome lattice in the STM topographic image is not attributable to the silicon atomic lattice, but originates from the local density of state (LDOS).

In order to reveal the detailed LDOS in the Kagome area, *dI/dV* tunneling spectra were collected from the sample at different sample biases. Spectra collected in the Kagome area display a broad peak at around 1.7 eV and a delta-function peak at 1.32 eV, which are not detectable from the bare $\sqrt{3} \times \sqrt{3}$ silicene surface, as shown in **Figure 2a**. The intensities of both peaks are extremely higher than for any features in STS on $\sqrt{3} \times \sqrt{3}$ silicene. Their intensities are also much higher than the interlayer-rotation-induced van Hove singularities (vHs) at -1.2 V and 0.75 V (*21*), as shown in the inset spectra in **Figure 2a**. We fitted both pronounced peaks by Lorentzian functions in order to reveal the origins of the corresponding electronic states, as shown in the **Figure 2b**. For the broad peak at about 1.7 V, two peak components with an interval of 0.13 V were revealed at 1.60 V and 1.73 V, respectively. These two components have similar spectral weights, band widths, and spectral shapes. In contrast, the delta-function peak at 1.32 eV can be well fitted by a single peak. Generally, such a pronounced delta-function peak in tunneling spectra is only observed in two scenarios. The first scenario involves the presence of isolated atoms, defects, or quantum dot systems because of their localized electrons (*18, 22*). At different sample biases, however, from -3 V to 3 V, we did not observe any topographic characteristics of isolated atoms, defects, or quantum dots in the STM images of the Kagome area. The first scenario can be therefore ruled out. The second scenario involves localization induced by quantum destructive interference. For instance, the electrons can be localized by destructive interference via multiple hopping paths (*6-8*), which is the core concept of quantum interference (*23*, *24*) in condensed matter physics. To explore the origin of these peaks, we conducted *dI/dV* mapping on the Kagome area at energies of 1.32 V and 1.7 V (**Figure 2d and 2e**), which correspond to the sharp peak and the broad peak, respectively. Coincidently, a clear electronic Kagome pattern is revealed by *dI/dV* mapping at the energy exactly corresponding to the peak at 1.32

V, as shown in **Figure 2d**. The results suggest that the sharp peak is associated with the frustrated Kagome geometry. Note that the delta-function peak with such a high intensity is not due to large tunneling matrix elements, because the intensity does not vary with different STM tips or other tunneling parameters in our tunneling spectra (**See Supplementary Note 7**).

In an electronic Kagome lattice, the behaviour of electrons can be understood by the schematic diagram in **Figure 1a**. At points A and C the wave functions are antiphase. If only electronic NN hopping exists, the hopping processes of A to B and C to B cancel each other out, which effectively localizes the states in the hexagons formed by A and C sites, as shown in **Figure 1a**. These localized states manifest themselves as a FB in momentum space or a peak in the density of states (DOS). In our STS mapping, there is always a similar "hexagonal-lattice" feature in the energy range from 1.60 V to 1.73 V, which is complementary to the Kagome lattice in real space (**Figure 2c and Supplementary Note 5**). This gives rise to two peaks in the *dI/dV* spectra. In contrast, the appearance of a Kagome pattern in the STS mapping at 1.32 V is indicative of the spatial distribution of the FB (*6-8*). The projection of the FB commonly induces a high-intensity delta-function peak in *dI/dV* spectra, because it is analogous to high degeneracy of electronic states. Based on the above discussion, it suggests that the pronounced peak at 1.32 V is attributable to the electronic localization induced by quantum destructive interference in Kagome geometry.

**Figure 3a** shows the topography of the Kagome lattice on the clean $\sqrt{3} \times \sqrt{3}$ silicene surface. Interestingly, in **Figure 3b**, the simultaneously acquired *dI/dV* mapping at 1.45 V reveals a strong density of states (DOS) near the edge between the Kagome area and the $\sqrt{3} \times \sqrt{3}$ silicene. As shown in **Figure 3c**, the topographical image of the border region reveals a ">" structural terrace edge of the Kagome area, as indicated by the white dashed line. Within the Kagome area, all the complete Kagome lattice cells closest to the terrace edge can be clearly identified. The A, B, and C sites in the complete Kagome lattice cell are labelled by blue, white, and red circles, respectively. When we connected their B sites, a boundary (marked by the black dashed line) which separates incomplete Kagome cells and complete Kagome cells can be seen and is defined as the "Kagome edge" hereafter. In STS mapping in the same region, the spatial distribution of the FB peak at 1.32 V is only confined within the area with complete Kagome lattices (left area to Kagome edge), but it is absent between the Kagome edge and the terrace edge (**Figure 3d and Supplementary Note 4**). The distance between the Kagome lattice and the terrace edge is about 2 nm in real space, which is significantly larger than the lattice constant of $\sqrt{3} \times \sqrt{3}$ silicene. Interesting, a ")"-shaped electronic state appears in the STS mapping with sample bias of 1.45 V, the spatial distribution of which matches perfectly with the Kagome edge (**Figure 3e**). This state spreads over several atomic rows with a width of about 2 nm in real space. It is much wider than the trivial edge state induced by edge dangling bonds or atomic reconstructions, which decays exponentially away from the terrace edge (*25, 26*). Indeed, there is a clear dangling-induced state along the terrace edge with a spatial distribution of less than 1 nm (**Figure 3f**). Therefore, the ")"-shaped electronic state at the sample bias of 1.45 V in the STS mapping is not attributed to dangling bonds but is given rise to Kagome edge. These mapping results suggest that electronic Kagome geometry indeed brings new electronic edge states due to symmetry broken in Kagome lattice.

Due to interlayer interaction in the twisted $\sqrt{3} \times \sqrt{3}$ silicene multilayer (*21*), the local potential around the AA stacking region should be strongly altered (*27*). These various local potential regions will act as strong scattering centres, leaving the remaining region as an effective periodical electronic lattice (*27*). Simple geometric structure analysis shows that this effective electronic lattice is a Kagome lattice (**See Supplementary Note 2**). Therefore, we can propose the hypothesis that, as a result of the twisting of $\sqrt{3} \times \sqrt{3}$ silicene layers, the periodic electronic potential of the system and electron hopping properties on certain atoms would change, which finally lead to the FB and the electronic Kagome lattice. As shown in **Supplementary Note 2**, the interlayer twisted $\sqrt{3} \times \sqrt{3}$ bilayer silicene with twisting angle of 21.8º induces a $\sqrt{21} \times \sqrt{21}$ superlattice. Due to the complexity of the large $\sqrt{21} \times \sqrt{21}$ unit cell, it is hard to construct an exact structural model of twisted $\sqrt{3} \times \sqrt{3}$ silicene multilayers. To test our hypothesis, we firstly used a $\sqrt{21} \times \sqrt{21}$ low-buckled silicene single layer with passivating hydrogen atoms on specific points (white ones in **Figure 4a**) to match the periodicity of the structural model proposed above and simulate the electron hopping between Si-layers. In the electronic band structure (**Figure 4b**), there is a FB above the Fermi level (**Figure 4c**), as well as the double-peak feature observed in the DOS (**Figure 4d**). The deviation from the FB near the Gamma point is induced by interlayer hopping. The simulated *dI/dV* mappings (**Figure 4c and 4d**) for the two bands clearly reproduce the Kagome lattice and hexagonal lattice features, which agree well with the experimental observations on *dI/dV* mappings (**Figure 2d and 2e**). All these results prove that our hypothesis is reasonable here. Moreover, when we used a pseudopotential with 0.5 H to simulate weaker interlayer electron hopping, the band near the Fermi level became much flatter (**see Figure S4 in Supplementary Note 2**). This result indicates that the formation of the FB should be highly dependent on the strength of the interlayer interaction. We also made a more realistic model for the simulation based on experimental observations. Two layers of $\sqrt{3} \times \sqrt{3}$ silicene are stacked with a twist of 21.8° (**Figure 4e**) between them resulting in a $\sqrt{21} \times \sqrt{21}$ unit cell, which is a composite moiré pattern with two moiré lattices. Therefore, an electronic Kagome lattice is induced by this composite moiré potential. Although the electronic band structure is very complex, and there are no clear features in the DOS (**Figure 4f**), one can still find two bands above 1.0 eV with dispersion close to the bands shown in **Figure 4b**. Also, the simulated *dI/dV* mappings (**Figure 4g and 4h**) are in good agreement with the experimental results.

It should note that the FB and a broad peak (BP) are located higher in energy than the vHs and the tail of the Dirac cone of $\sqrt{3} \times \sqrt{3}$ silicene (*21, 28-30*). This phenomenon is close to the situation for nearly free electron states or the image potential states observed in carbon-based systems (*31-33*). The tail of the Dirac cone in graphene is about 3.0 eV, so those states can only be observed in the high energy range (*31-32*). Here, the free electron states have originated from the interlayer twisting of silicene. With the twisting between the Si-layers, a screened periodic Coulomb potential field is formed. Then, the nearly free electron states are induced, leading to the observation of an electronic Kagome lattice and the FB at 1.3 eV. Furthermore, similar to the observed delta-function FB peak, the free electron states always show strong peaks, with the *dI/dV* signal stronger than other features (*31, 33*). This scenario is also supported by the strong edge state at an energy right above the FB energy (**Figure 2d**) at the Kagome edge, which represents the scattering of nearly free electrons on the boundary of the screened periodic Coulomb potential field. This is an inevitable result of broken symmetry in this Kagome system, just like the surface states in semiconductors are always at an energy right above their valence bands. Moreover, on a 2D $\sqrt{3} \times \sqrt{3}$ silicon film grown on a metal substrate, ferromagnetism with a transition temperature higher than 170 K is predicted (*34*). This means that the electronic Kagome lattice observed on multilayer silicene may meet the critical preconditions (Kagome lattice, FB, and magnetism) for exotic topological phenomena (*6-8*), making the observed edge state a promising candidate.

In summary, we experimentally realized an electronic Kagome lattice with the emergence of an electronic FB on a twisted multilayer silicene surface. The local potential modulation attributed to the interlayer interaction gives rise to generation of electronic Kagome lattice, and thereby, electronic FB. The electronic edge state induced by symmetry broken in Kagome regimes has been observed. The present experimental results for the band structure and edge states are consistent with the theoretically predicted topological properties of the Kagome lattice.

**Acknowledgements:** This work was supported by the Australian Research Council (ARC) through Discovery Projects (DP140102581, DP160102627 and DP170101467) and Linkage Infrastructure, Equipment and Facilities (LIEF) grants (LE100100081 and LE110100099). Z. L. would like to acknowledge support by the University of Wollongong through the Vice Chancellor's Postdoctoral Research Fellowship Scheme. Z. H. would like to acknowledge China Scholarship Council (CSC) for the financial support (No.201506200150).The work was partially supported by the NSF of China (grants nos. 11334011, 11322431, 11304368, 11674366, 11674368, 21203099), Fok Ying Tung Education Foundation (No. 151008), and the Strategic Priority Research Program of the Chinese Academy of Sciences (grant no. XDB07020100). The authors thank Dr. T. Silver for her valuable comments on this work. The authors also thank Xiao-Gang Wen (Massachusetts Institute of Technology), Feng Liu (University of Utah), and Zheng Liu (Tsinghua University) for the valuable discussions on this work.

**Author Contributions**

Z. L., J. Z., and Y. D. designed the experimental plan. Z. L. and L. C. prepared the samples. Z. L., J. Z., H. F. and Y. D. did STM characterizations. Z. H. carried out modelling and DFT calculations. L. W., X. Xu., X. W., C. Z., K. W., and S. X. D helped with data analysis. All authors participated in discussion on the data. Z. L., Z. H. and Y. D wrote the paper.

**Additional information:** The authors declare no competing financial interests.

**Figures and figure captions**

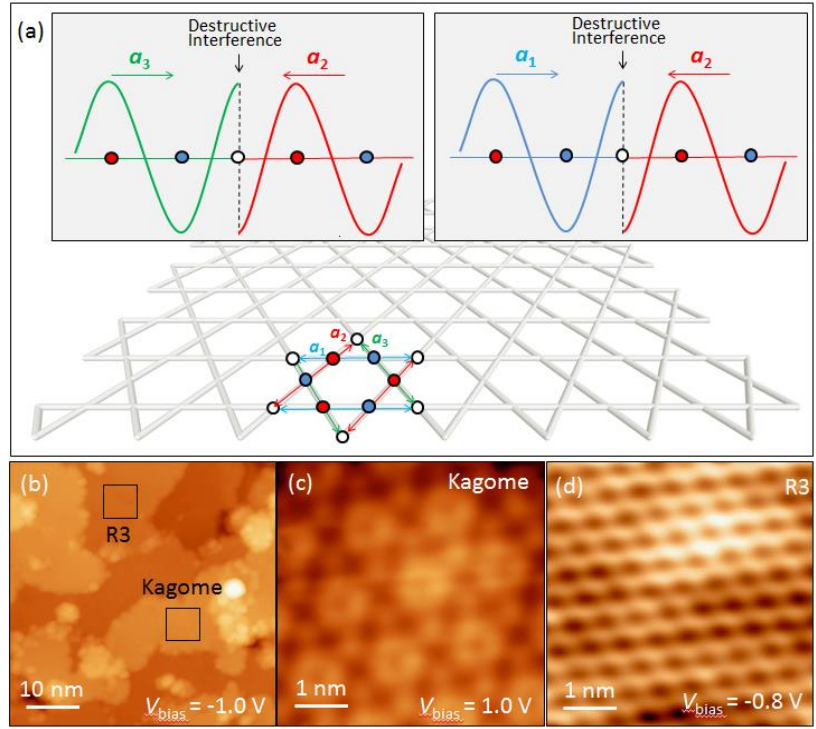

**Figure 1. Kagome lattice imaged by STM. a**, Schematic diagram of destructive quantum interference inducing a FB in the Kagome lattice. Three different sites: A, B, and C are marked with three different colours: blue, white, and red, respectively. The A and C sites have same wave amplitude, but the A site is antiphase with the C site. Considering only NN hopping, the possibility of an electron escaping the hexagon is determined by overlapping two hopping vectors, hopping from A to B and C to B. Because the A and C sites have same amplitude and same sign, the two hopping wave vectors will have same amplitude and opposite sign, as shown in the schematic drawings. Thus, these two vectors will cancel each other out and lead to zero possibility of an electron hopping from the hexagon to the B site, which means that the electrons are localized in the hexagons in Kagome lattice. In momentum space, the localized electron states mean infinite effective mass and that the energy band is flat. **b**, STM topographical image (image size 50 nm × 50 nm, sample bias $V_s$ = 2.7 V, tunneling current $I_t$ = 100 pA) of multilayer silicon grown on Ag(111) substrate. **c**, STM topographical image (5 nm × 5 nm, $V_s$ = 1 V, $I_t$ = 100 pA) giving an enlarged view of the lower right black rectangle in **b** in the Kagome area, with a Kagome lattice constant of 1.7 nm. **d**, STM topographical image (5 nm × 5 nm, $V_s$ = -0.8 V, $I_t$ = 100 pA) giving an enlarged view of the upper left black rectangle in **b**. It shows a honeycomb lattice with a lattice constant of 0.64 nm, which is ascribed to the $\sqrt{3} \times \sqrt{3}$ phase of silicene.

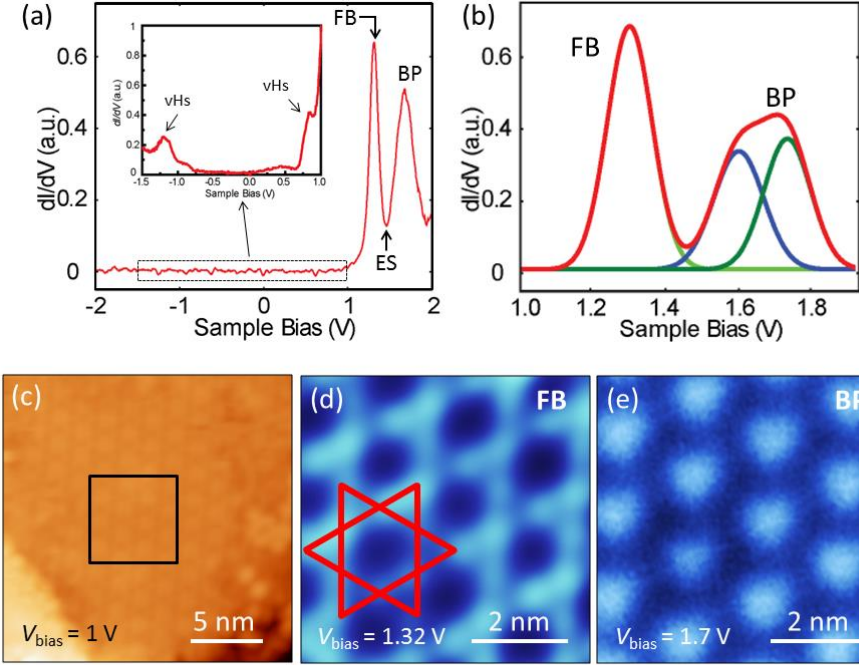

**Figure 2. Electronic structure of Kagome area. a**, STS spectrum ($V_s$ = 2 V, $I_t$ = 100 pA) from the Kagome area shows two distinctive peaks: a delta-function peak at 1.32 V (flat band peak, FB) and a broad peak (BP) centred around 1.7 V. The valley position is corresponding to the energy of Kagome edge state (ES). The inset figure shows the STS spectrum from -1.5 V to 1 V. Two van Hove singularities (vHs) peaks are clearly resolved at -1.2 V and 0.75 V. **b**, The fitting results for the delta-function peak and the broad peak. The delta-function peak is fitted excellently by a single delta-function peak. The broad peak is fitted by Lorentzian functions. The two peak components are centred at 1.60 V and 1.73 V. **c**, STM image of large scale Kagome lattice (20 nm × 20 nm, $V_s$ = 1 V, $I_t$ = 100 pA). **d** and **e**, DOS mappings of the region enclosed by the black square in **c** at the FB (6 nm × 6 nm, $V_s$ = 1.32 V, $I_t$ = 100 pA) and BP (6 nm × 6 nm, $V_s$ = 1.7 V, $I_t$ = 100 pA) energies. The Star of David is a guide to the eye.

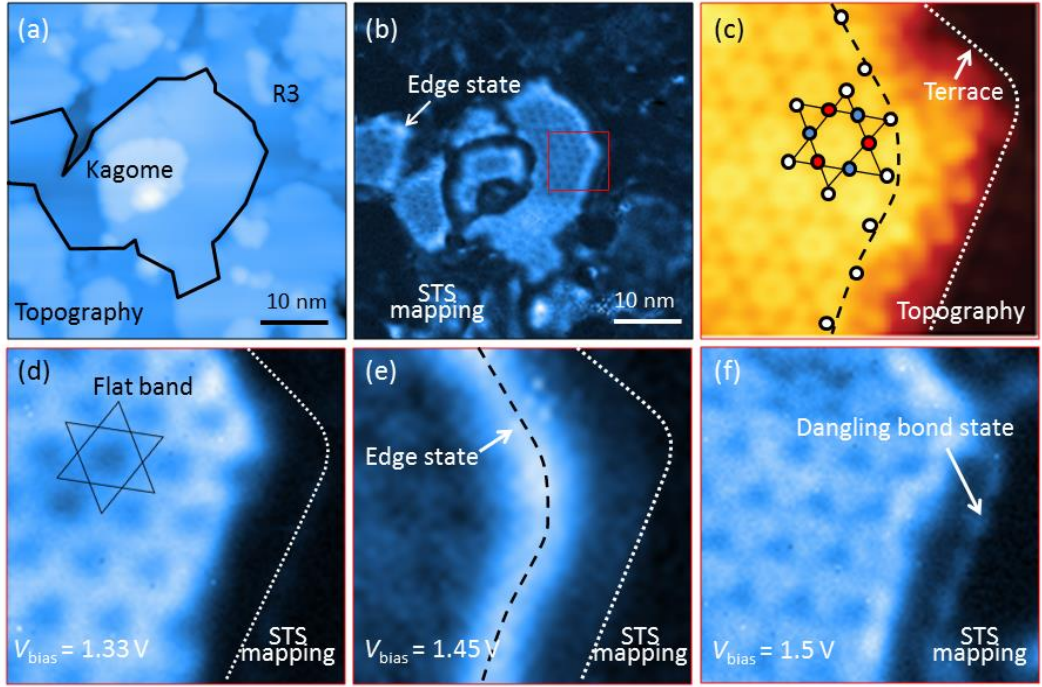

**Figure 3. Electronic structure of Kagome area. a,** Large-area image of Kagome lattice surrounded by √3 × √3 silicene (R3 area) (50 nm × 50 nm, $V_s$ = 3 V, $I_t$ = 100 pA). The boundary between them is marked by the black solid line. **b,** DOS mapping simultaneous obtained at 1.45 V, which is the corresponding edge state energy. The position of the edge state is highlighted by the white arrow. **c,** Topographical image (10 nm × 10 nm, $V_s$ = 1 V, $I_t$ = 100 pA) of the border region between the Kagome lattice (upper left) and √3 × √3 silicene (lower right), which is enclosed by the red square in **b**. The Kagome lattice is constructed from three different sites, which are marked by blue, white, and red circles. The Kagome edge is marked by the black dashed line connecting the white circles. The terrace edge is marked by the white dashed line. **d** and **e**, DOS mapping of the FB energy (10 nm × 10 nm, $V_s$ = 1.33 V, $I_t$ = 100 pA) clearly reveals the Kagome pattern on the left of these images. The Star of David is a guide to the eye. **e**, DOS mapping of the edge state energy (10 nm × 10 nm, $V_s$ = 1.45 V, $I_t$ = 100 pA) shows higher DOS along the dashed line in **c**. **f**, Dangling bonds are revealed by DOS mapping (10 nm × 10 nm, $V_s$ = 1.5 V, $I_t$ = 100 pA).

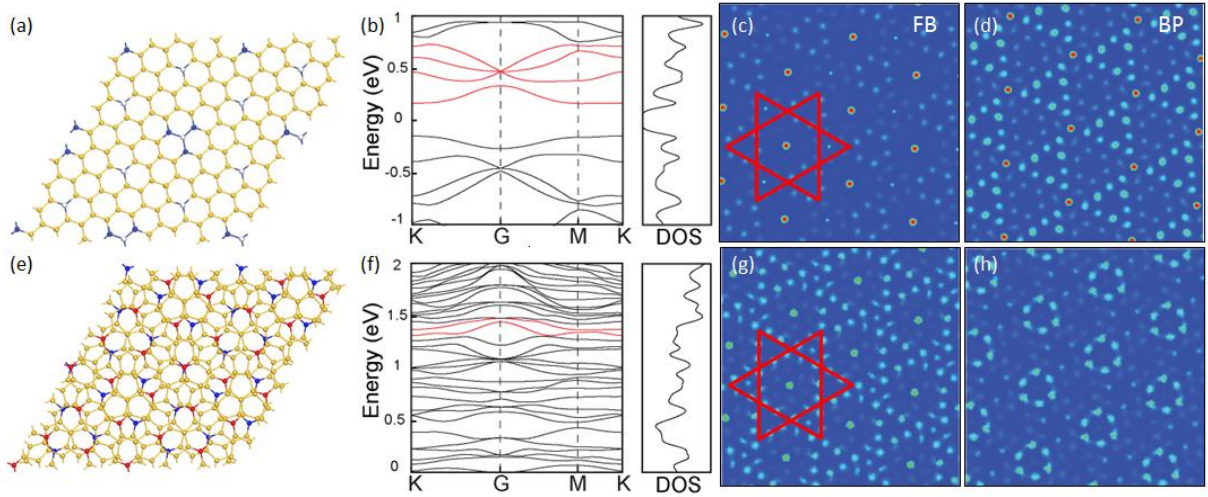

**Figure 4. DFT simulation of electronic Kagome lattice. a**, The model of low buckled single layer silicene passivated by hydrogen atoms (marked in white) for simulating the periodically change on local potential. The hydrogen passivated silicon atoms are marked in blue. **b**, Band structure of the model (in **a**) from DFT calculations shows a FB around 0.2 eV above the Fermi surface and a broad band with higher energy (both marked in red). The right part of **b** shows the DOS. **c** and **d**, Local density of states (LDOS) of the FB and broad band shows a clear Kagome lattice and hexagonal lattice respectively. The red Star of David is a guide to the eyes. **e**, Structure model of two layers of $\sqrt{3} \times \sqrt{3}$ silicene twisted by 21.8°. The $\sqrt{3} \times \sqrt{3}$ buckled atoms in the top layer and bottom layer are marked in red and blue, respectively. As with **b**, band structure of this model (in **f**) also shows a FB around 1.30 eV and a broad band with slightly higher energy (both marked in red). In this model, the LDOS of the FB (in **g**) and broad band (in **h**) again clearly shows the Kagome lattice and hexagonal lattice, respectively.